\documentclass{article}

\def\dg{\dagger}   \def\vphi{\varphi}  \def\ra{\rangle}    \def\la{\langle}
\def\alf{\alpha}

\def\fns{\footnotesize}   \def\beq{\begin{equation}} \def\eeq{\end{equation}}
\def\bea{\begin{eqnarray}}  \def\eea{\end{eqnarray}}
\def\tld{\tilde}             \def\lb{\label}  \def\dstyle{\displaystyle}
\def\bt{\begin{tabular}}     \def\et{\end{tabular}}

\begin{document}

\title{Pseudo-Boson Coherent and Fock States}

\author{D.A. Trifonov \footnote{e-mail: dtrif@inrne.bas.bg}\\
{\normalsize Institute for Nuclear Research and Nuclear Energy}\\
{\normalsize Bulgarian Academy of Sciences} \\
{\normalsize 72 Tzarigradsko chausse Blv, 1784 Sofia, Bulgaria}}

\maketitle
\vspace{-6cm}
e-print arXiv: quant-ph/0902.3744 
\vspace{6cm}

\begin{abstract}
Coherent states (CS) for non-Hermitian systems are introduced as eigenstates of pseudo-Hermitian boson annihilation operators. The set of these CS includes two subsets which form bi-normalized and bi-overcomplete system of states. The subsets consist of eigenstates of two complementary lowering pseudo-Hermitian boson operators.

Explicit constructions are provided on the example of one-parameter family of pseudo-boson ladder operators. The wave functions of the eigenstates of the two complementary number operators, which form a bi-orthonormal system of Fock states, are found to be proportional to new polynomials, that are bi-orthogonal and can be regarded as a generalization of standard Hermite polynomials.
\end{abstract}



\section{Introduction}

In the last decade  a growing interest is shown in the non-Hermitian {\cal PT}-symmetric (or pseudo-Hermitian) quantum mechanics. For a review with an enlarged list of references see the recent papers \cite{Bender1, Dorey1}. This trend of interest was triggered by the papers of Bender and coauthors \cite{Bender2}, where the Bessis conjecture about the reality and positivity of the spectrum of Hamiltonian $H= p^2+x^2+ix^3$ was proven ('using extensive numerical and asymptotic studies') and argued that the reality of the spectrum is due to the unbroken $PT$-symmetry. The Bessis--Zinn Justin conjecture about the reality of the spectrum of the $PT$-symmetric Hamiltonian $p^2 - (ix)^{2\nu}$ for $\nu\geq 1$ has been proven in Ref. \cite{Dorey2}. A criterion for the reality of the spectrum of non-Hermitian $PT$-symmetric Hamiltonians is provided in Ref. \cite{Caliceti}.
Mustafazadeh \cite{M-zadeh1} has soon noted that all the $PT$-symmetric non-Hermitian Hamiltonians studied in the literature belong to the class of pseudo-Hermitian Hamiltonians. A Hamiltonian $H$ was said to be pseudo-Hermitian if it obeys the relation \cite{M-zadeh1}
\beq\label{H eta}
H^{\#} := \eta^{-1}H^\dg\eta = H,
\eeq
where $\eta$ is an invertible Hermitian operator. $H^{\#}$ was called {\it $\eta$-pseudo-Hermitian conjugate} to $H$, shortly {\it $\eta$-pseudo-adjoint } to $H$. The $PT$-symmetric Hamiltonian $H= p^2-(ix)^{2\nu}$, examined in Refs. \cite{Bender2, Dorey2}, obeys (\ref{H eta}) with $\eta$ equal to the parity operator $P$. The spectrum of a diagonalizable pseudo-Hermitian $H$ is either real or comes in complex conjugate pairs. A diagonalizable (non-Herimitian) Hamiltonian $H$ has a real spectrum iff it is pseudo-Hermitian with respect to positive definite $\eta$ \cite{M-zadeh2}. In terms of $P$ and $T$ operations the reality of the spectrum of $H$ occurs if the $PT$ symmetry is exact (not spontaneously broken) (see e.g. Refs. \cite{Bender1, Dorey1} and references therein). In fact many of the later developments in the field are anticipated in the paper by Scholtz et al \cite{Scholtz} (see comments in Ref.  \cite{Bender1}).

\smallskip

In the present paper we address the problem of construction of pseudo-Hermitian boson (shortly pseudo-boson) creation and annihilation operators and related Fock states and coherent states (CS). For pseudo-fermion system ladder operator CS have been introduced by Cherbal et al \cite{Cherbal} on the example of two-level atom interacting with a monochromatic em field in the presence of level decays.  For non-Hermitian $PT$-symmetric system CS of Gazeau-–Klauder type have been constructed, on the example of Scarf potential, by Roy et al \cite{Roy}.  Annihilation and creation operators in non-Hermitian (supersymmetric) quantum mechanics were considered by Znojil \cite{Znojil1}. For the bosonic $PT$ symmetric singular oscillator (which depicts a double series of real energy eigenvalues) \cite{Znojil2} ladder operators and eigenstates of the annihilation operators have been built up by Bagchi and Quesne \cite{Quesne}.  Our main aim here is the construction of overcomplete families (in fact bi-overcomplete) of ladder operator CS for pseudo-bosons. The problem of pseudo-boson ladder operators is considered in section 2. In the third section we consider the construction of eigenstates of pseudo-boson number operators. The pseudo-boson CS are introduced and discussed in section 4. Explicit example of pseudo-boson ladder operators and related eigenstates is provided and briefly commented in section 5. Outlook over the main results is given in the Conclusion.

\section{Pseudo-boson ladder operators }

With the aim to construct pseudo-Hermitian boson (shortly {\it pseudo-boson}) coherent states (CS) we address the problem of ladder operators that are {\it pseudo-adjoint} to each other.
In analogy with the pseudo-Hermitian fermion (phermion) annihilation and creation operators \cite{M-zadeh3} the $\eta$-pseudo-boson ladder operators $b,\,\, b^\#:=\eta^{-1}b^\dg\eta$  can be defined by means of the commutation relation
\beq\label{pb}
 [b,b^\#] \equiv b b^\# - b^\# b =1.
\eeq
If $\eta=1$ the standard boson operators $a,\, a^\dg $ are recovered.

From (\ref{pb}) it follows that the pseudo-Hermitian (pseudo-selfadjoint) operator $b^\# b \equiv N$  commutes with $b$ and $b^\#$  according to
\beq\label{Nb}
 [b,N] =  b,\quad [b^\#,N] = -b^\# ,
\eeq
and could be regarded as {\it pseudo-boson  number operator}. For a pair of non-Hermitian operators $b,\,\tld b$ with commutator $[b,\tld b]=1$, the existence of $\eta$ such that $\tld b = \eta^{-1}b^\dg \eta \equiv b^\#$, stems from the existence of $b$-vacuum. We have the following \smallskip

{\bf Proposition 1.} If the operators $b$ and ${\tld b}$ and a state $|0\ra$ satisfy 
 \begin{equation}\label{b|0>} 
[b,\tld b] = 1,\quad b|0\ra = 0, 
\end{equation}
 then $\tilde b$ is $\eta$-adjoint to $b$ with 
\begin{equation}
\eta = \sum_{n=0}|\vphi_n\ra\la\vphi_n| ,
\end{equation}
 where $|\vphi_n\ra$ are eigenstates of $N'=b^\dg \tld{b}^\dg$.
\smallskip

{\bf Proof.} Note first that the non-Hermitian operator $N = \tld b b$ is diagonalizable and with real and discrete spectrums. It's eigenstates can be constructed acting on the $b$-vacuum by the operators $\tld b$ correspondingly (see the next section). The spectrum of a diagonalizable non-Hermitian operator $H$ is real iff $H$ is $\eta$-pseudo-Hermitian (theorem of Ref.   \cite{M-zadeh1}), and this $\eta$ may be chosen as a sum of projectors onto the eigenstates of $H^\dg$ \cite{M-zadeh1}. In our cases $H=N$ and $H^\dg = N^\dg = b^\dg\tld b^\dg$. This ends the proof of the Proposition 1.
\smallskip

For the sake of completeness however we provide in the next section the construction (and brief discussion) of the eigenstates of $N$, $N'$.
\smallskip

{\bf Remark.} A similar Proposition can be formulated and proved for a pair of non-Hermitian nilpotent operators $g, \,\,\tld g$, $g^2=0$ with anticommutator $\{g,\tld g\}=1$: just replace $b$, $\tld b$, $[b,\tld b]$ with $g$, $\tld g$, $\{g,\tld g\}$ .

\section{Pseudo-boson Fock states}

The eigenstates of pseudo-boson number operators  $N = \tld b b$ can be constructed acting on the ground states $|0\ra$ by the raising operator $\tld b$:
\bea \label{psi_n}
 |\psi_n\ra  =   \frac{1}{\sqrt{n!}}{\tilde b}\,^n|0\ra , \\  
 N |\psi_n\ra  =  \tld b b |\psi_n\ra = n|\psi_n\ra.
\eea
However in view of $\tilde b \neq  b^\dg$ these number states are not orthogonal.
It turned out that a {\it complementary pair} of pseudo-boson ladder operators and number operator exist, such that the system of the two complementary sets of number states form the so-called {\it bi-orthogonal and bi-complete sets}. Indeed, if $\tilde b$ is creating operator related to $b$, then, on the symmetry ground, we could look for new operators $b'$ for which $b^\dg$ is the creating operator,
\begin{equation}\label{b'}
[ b', b^\dg] =1 .
\end{equation}
The pairs of "prime"-ladder operators $b',\,b'^\dg$ is just $\tilde b^\dg,\,b^\dg$, and the "prime" number operator is
\beq\lb{N'}
 N'= b'^\dg  b' =  b^\dg \tld b^\dg .
\eeq
The eigenstates of $N'$ are constructed in a similar way acting with $b^\dg$ on the
$b'$-vacuums $|0\ra'$ :
\begin{equation}\label{phi_n}
|\vphi_n\ra = \frac{1}{\sqrt{n!}}( b^\dg)^n|0\ra'.
\end{equation}
The existence of the $b'$-vacuum $|0\ra'=|\vphi_0\ra$ follows from the properties of the pseudo-Hermitian operators $H$ with real spectra \cite{M-zadeh1, M-zadeh2}: the spectrum of $H$ and $H^\dg$ coincide since they are related via a similarity transformation. In our case $H=\tld b b$, $H^\dg = b^\dg \tld b^\dg \equiv b^\dg b'$.

Using the commutation relations of the above described ladder operators one can easily check that if $\la 0|0\ra' = 1$ then the "prime" number-states $|\vphi_n\ra$ are {\it bi-orthonormalized} to $|\psi_n\ra$ (that is $\la\psi_n | \vphi_m\ra = \delta_{nm}$),
and form together the {\it bi-complete} system of states $\{|\psi_n\ra, |\vphi_n\ra\}$:
\begin{equation}\label{1}
\sum_n|\psi\!_n\ra \,\la\vphi_n| = 1 = \sum_n|\vphi\!_n\ra \,\la\psi_n| .
\end{equation}

The set $\{|\psi_n\ra,|\vphi_n\ra\}$ can be called the set of Fock states for pseudo-Hermitian boson system (shortly {\it pseudo Fock states}).
In terms of the projectors on these states the pseudo-boson ladder operators $ b,\,\,\tilde b$ can be represented as follows
\begin{equation}\label{b,b'}
\begin{tabular}{ll}
$  b = \sum_n \sqrt{n}|\psi_{n-1}\ra\,\la\vphi_n|,\quad
\tilde b = \sum_n \sqrt{n}|\psi\!_n\ra \,\la\vphi_{n-1}|$,\\[2mm]
$  b' = \sum_n \sqrt{n}|\vphi_{n-1}\ra\,\la\psi_n|,\quad
 b^\dg = \sum_n \sqrt{n}|\vphi\!_n\ra \,\la\psi_{n-1}|$.\\[2mm]
\end{tabular}
\end{equation}

Now consider the operator \cite{M-zadeh1}
\begin{equation}\label{eta}
 \eta = \sum_n |\vphi_{n}\ra\,\la\vphi_n|.
\end{equation}
This is Hermitian, positive and invertible operator, 
$\eta^{-1} = \sum_n |\psi_{n}\ra\, \la\psi_n|$ .
From the above expressions of $\eta$ and $\eta^{-1}$ one can see that $|\vphi_n\ra = \eta |\psi_n\ra$.

Finally one can easily check (using (\ref{b,b'}) ) and (\ref{eta}) that  $\tld b$ is $\eta$-pseudo-adjoint to $b$,\,\, $b'$ is $\eta^{-1}$-pseudo-adjoint to $b^\dg$,
\begin{equation}
\tilde b =  \eta^{-1} b^\dg \eta, \quad  b' = \eta b \eta^{-1},
\end{equation}
and $N$ and $N'$ are $\eta$- and $\eta^{-1}$ pseudo-Hermitian:  $N^\# := \eta^{-1} N^\dg \eta = N$,  $(N')^\# := (\eta')^{-1} (N')^\dg \eta' = N'$, $\eta'=\eta^{-1}$.

\section{Pseudo-boson coherent states}\label{sec:3}

We define coherent states (CS) for the pseudo-Hermitian boson systems as eigenstates of the corresponding pseudo-boson annihilation operators. In this aim we 
introduce the pseudo-unitary displacement operator $D(\alf) = \exp(\alf  b^\# - \alf^* b)$ and construct eigenstates of $b$ as displaced ground state $|0\ra$,
\begin{equation}\label{|alf>}
 b |\alf\ra = \alf |\alf\ra,  \quad \rightarrow \quad |\alf\ra = D(\alf)|0\ra,
\end{equation}
where $\alf \in C$. Using BCS formula one gets the expansion
\begin{equation}
|\alf\ra = e^{-\frac{1}{2}|\alf|^2} \sum_n \frac{\alf^n}{\sqrt{n!}}|\psi_n\ra .
\end{equation}
The structures of the above two formulas are the same as those for the case Hermitian boson CS (the Glauber canonical CS \cite{Glauber}), but the properties of our $D(\alf)$ and  $|\alf\ra$ are different. First note that $D(\alf)$ is not unitary. Therefore
$|\alf\ra$ are not normalized. Second, the set $\{|\alf\ra; \alf \in C\}$ is not overcomplete (since $|\psi_n\ra$ are not orthogonal).

The way out of this impasse is to consider the eigenstates of the dual ladder operator $ b'=( b^\#)^\dg$, which take the analogous to (\ref{|alf>}) form,
\begin{equation}\label{|alf>'}
 b' |\alf\ra' = \alf |\alf\ra', \quad  \rightarrow \quad |\alf\ra' = D'(\alf)|0\ra',
\end{equation}
where $D'(\alf) = \exp(\alf  b^\dg - \alf^* b')$ is the complementary pseudo-unitary displacement operator. Therefore the eigenstates $|\alf\ra'$ are again non-normalized and do not form overcomplete set. However they are {\it bi-normalized} to $|\alf\ra$ (that is $\la \alf|\alf\ra' = 1$) and the system $\{|\alf\ra, |\alf\ra'; \alf \in C\}$ is {\it bi-overcomplete} in the following sense
\begin{equation}\label{overcompl}
\frac{1}{\pi} \int d^2\alf \, |\alf\ra'\,\,\la \alf| = 1,\quad  \frac{1}{\pi} \int d^2\alf  \, |\alf\ra\,\,'\!\la \alf| = 1.
\end{equation}
It is this bi-overcomplete set that we call {\it pseudo-boson CS}, or CS of pseudo-boson systems. More precisely they are $\eta$- pseudo-boson CS. When $ \eta=1$ these states recover the famous Glauber CS $|\alf\ra=\exp(\alf a^\dg - \alf^* a)|0\ra$, where $a,\,a^\dg$ are canonical boson annihilation and creation operators.

\section{Example}

In this section we illustrate the above described scheme of construction of pseudo-boson Fock states and CS on the example of the following one-parameter family of non-Hermitian operators,
\begin{equation}\label{b(s)}
\begin{tabular}{l}
$ b(s) = a + s a^\dg$,\\
$\tilde b(s)  =   s a  +(1+s^2) a^\dg,$
\end{tabular}  \end{equation}
where $s\in (-1,1)$ and $a,\,a^\dg$ are Bose annihilation and creation operators:  $[a,a^\dg]=1$.
It is clear that $ b(0)= a$, $\tilde b(0)= a^\dg$ and $\tilde b(s) \neq  b^\dg(s)$. In this way the parameter $s$ could be viewed as a measure of deviation of $b(s)$ and $b^\#(s)$ from the canonical boson operators $a$ and $a^\dg$.

\subsection{Pseudo-boson Fock state wave functions}

The $b(s)$- and $b'$-vacuums $|0\ra$ and $|0\ra'$ do exist.
Using the coordinate representation of $b(s)$ and $\tld b(s)$,
\bea\lb{b(s)2}
 b(s) = \frac{1}{\sqrt{2}}\left( (1+s)x - (1-s)\frac{d}{dx} \right),\\
 \tld b(s) = \frac{1}{\sqrt{2}}\left( (s +1 +s^2)x - (s-1-s^2)\frac{d}{dx} \right), \lb{tld b(s)2}
\eea
we find the wave functions of $|0\ra$ and $|0\ra'$ ($ {\cal N}(s) = \left(\pi (1-s)(1-s+s^2)\right)^{-\frac{1}{4}}$):
\begin{equation}\label{psi_0}
\psi_0(x,s) = {\cal N}(s)\exp\left(-\frac{1+s}{2(1-s)}x^2\right) ,
\end{equation}
\beq \lb{vphi_0}
\vphi_0(x,s) = {\cal N}(s)\exp\left(- \frac{1+s+s^2}{2(1+s^2-s)}x^2\right),
\eeq
The above two wave functions are bi-normalized, that is $\la\vphi_0|\psi_0\ra = 1$, if the parameter $s$ is restricted in the interval $(-1,1)$, that is $-1 < s <1 $.

Therefore, according to Proposition 1 (and the related development in the previous section)  $\tilde b(s)$ is $\eta$-pseudo-adjoint to $b(s)$ and (for $s^2 <1$) the bi-orthonormalized Fock states and bi-overcomplete CS can be explicitly realized.  The wave functions of the pseudo-boson Fock states (\ref{psi_n}) and (\ref{phi_n}) are obtained in the following form:
\begin{equation}\label{psi_n 2}
\begin{tabular}{l}
$\dstyle \psi_n(x,s) =  \frac{1}{\sqrt{2^n n!}} P_n(x,s) \psi_0(x,s),$\\[3mm]
$\dstyle \vphi_n(x,s) = \frac{1}{\sqrt{2^n n!}} Q_n(x,s) \vphi_0(x,s)$,
\end{tabular}
\end{equation}
where $P_n(x,s)$, $Q_n(x,s)$ are polynomials of degree $n$ in $x$, defined by means of the following recurrence relations
\begin{equation}\label{PQ}
\begin{tabular}{l}
$\dstyle P_n = \frac{2}{1-s}xP_{n-1} + (n-1) \frac{2(s-s^2-1)}{1-s}P_{n-2}$,\\[3mm]
$\dstyle Q_n = \frac{2}{1+s^2-s}xQ_{n-1} + (n-1) \frac{2(s-1)}{1+s^2-s} Q_{n-2}$.
\end{tabular}
\end{equation}
For the first three values of $n$, $n=0,1,2$, the polynomials $P_n(x,s)$ and $Q_n(x,s)$ read:
\begin{equation} \nonumber 
\begin{tabular}{l}
$\dstyle P_0 = 1,\quad P_1= \frac{2}{1-s}x,\qquad P_2 =  \frac{4}{(1-s)^2}x^2 + \frac{2(s-s^2-1)}{1-s}$, \\ [3mm]
$\dstyle Q_0 = 1,\quad Q_1= \frac{2}{1-s+s^2}x,\,\,\, Q_2= \frac{4}{(1-s+s^2)^2} x^2 + \frac{2(s-1)}{1-s+s^2}$.
\end{tabular}
\end{equation}
At $s=0$ these two polynomials $P_n(x,0)$ and $Q_n(x,0)$ coincide and recover the known Hermite polynomials $H_n(x)$. Therefore $P_n(x,s)$ and $Q_n(x,s)$ can be viewed as two different generalizations of $H_n(x)$. They are not orthogonal. Instead of the orthogonality they satisfy the {\it bi-orthonality} relations, the weight function being $w(x,s) = \psi_0(x,s)\vphi_0(x,s)$,
\begin{equation}\label{int PQ}
\int  P_n(x,s)Q_m(x,s)\, \psi_0(x)\vphi_0(x)  = n!2^n \delta_{nm}.
\end{equation}
Therefore $P_n(x,s)$ and $Q_n(x,s)$ realize {\it bi-orthogonal generalization} of $H_n(x)$.
The relations between these three types of polynomials may be illustrated by the following diagram (where $\mu(s) = 1/(1-s)(1-s+s^2)$):
\vspace{-10mm}

\unitlength 0.75mm
\linethickness{0.6pt}
\begin{picture}(21.00,40.00)
\put(17.11,15.16){\vector(-1,-2){8.00}} 
\put(17.11,15.16){\vector(1,-2){8.00}} 
\put(12.0,-5.0){\makebox(0,0)[cc]{$P_n(x)$}}
\put(30.0,-5.0){\makebox(0,0)[cc]{$Q_n(x)$}}
\put(22.0,17.0){\makebox(0,0)[cc]{$H_n(x)$}}
\put(40.00,17.0){\vector(1,0){8}} 

\put(55.00,17.0){\makebox(0,0)[lc]{\footnotesize $\int H_n H_m  e^{-x^2}dx = \sqrt{\pi}2^n n!\,\delta_{nm}$}}
\put(40.00,-5.0){\vector(1,0){8}} 
\put(55.00,-5.0){\makebox(0,0)[lc]{\footnotesize $\int P_nQ_me^{-\mu(s)x^2}dx = {\cal N}^{-2}(s)2^n n!\, \delta_{nm}$}}
\end{picture}
\vspace{10mm}

It is worth emphasizing that the above bi-orthogonal generalization of $H_n(x)$ is not unique. If instead of (\ref{b(s)}), we take another pair of operators satisfying the Proposition 1, say
\begin{equation}\label{b(s)3}
\begin{tabular}{l}
$ b_2(s) = a + s a^\dg$,\\
$\tilde b_2(s)  =   -s a  +(1-s^2) a^\dg,$
\end{tabular}  \end{equation}
and apply the above described scheme, we would get another similar pair of bi-orthogonal polynomials.
\medskip

\subsection{Pseudo-boson CS wave functions}

In coordinate representation equations (\ref{|alf>}), (\ref{|alf>'}) for the eigenstates of $b(s)$ and $b'(s)={b^\#}^\dg(s)$ ($b(s)$ being given in (\ref{b(s)2})) lead to the following wave functions for any $\alf\in C$,
\beq\lb{psi_alf}
\psi_\alf(x,s) = {\cal N}(s,\alf)\exp\left[-\frac{1+s}{2(1-s)}x^2 + \frac{\sqrt{2}\alf}{1-s}x \right],
\eeq
\beq\lb{phi_alf}
\vphi_\alf(x,s) = {\cal N}'(s,\alf)\exp\left[-\frac{1+s+s^2}{2(1-s+s^2)}x^2 + \frac{\sqrt{2}\alf}{1-s+s^2}x\right],
\eeq
where ${\cal N}$ and ${\cal N}'$ are bi-normalization constants. Up to constant phase factors they are determined by the bi-normalization condition $\la \psi_\alf|\vphi_\alf\ra = 1$. We put $\alf = \alf_1 + i\alf_2$ and find
\beq\lb{nn'}
\bt{l}
$(N^*N')^{-1} = \int \exp\left[-\frac{x^2}{(1-s)(1-s+s^2)} + \sqrt{2}x\frac{ (2-2s+s^2)\alf_1 + i\alf_2 s^2}{(1-s)(1-s+s^2)} \right]dx $ \\[3mm]
$ \hspace{7mm} = \sqrt{\pi (1-s)(1-s+s^2)} \exp\left[\frac{\left( 2 \alf_1  -2\alf_1s + \alf_1 s^2  + i\alf_2 s^2 \right)^2}{2(1-s)(1-s+s^2)}\right]$.
\et \eeq
At $s=0$ we get, up to a constant phase factor, ${\cal N}= {\cal N}' = (1/\pi^{1/4})\exp(-\alf_1^2)$. At $s=0$ both $\psi_\alf(x,0)$ and $\vphi_\alf(x,0)$ recover the wave functions of Glauber canonical CS.

The constructed Fock states and CS are time independent, and can be used as initial states (initial conditions) of pseudo-boson systems. Important question arises of the {\it temporal stability } of these states. In analogy with the case of Hermitian mechanics we define temporal stability of a given set of states by the requirement the time evolved states to belong to the same set. This means that, up to a time-dependent phase factor, the time-dependence of any state from the initial set should be included in the time-dependent parameters only. Clearly the time dependent parameters should remain in the same domain as defined initially.

For our CS the temporal stability means that the time evolved wave functions $\psi_\alf(x,s,t)$, $\vphi_\alf(x,s,t)$ should keep the form
\beq\lb{t}
\bt{l}
$\dstyle \psi_\alf(x,s,t) = e^{i\chi(t)}\psi_{\alf(t)}(x,s)$,\\[2mm]
$\dstyle \vphi_\alf(x,s,t) = e^{i\chi'(t)}\vphi_{\alf(t)}(x,s))$,
\et
\eeq
where $\chi(t), \chi'(t)\in R$, $\alf(t)\in C$ and $s(t)^2<1$. It is clear, that if the time evolved CS obey (\ref{t}) they remain eigenstates of the same ladder operators $b(s)$ and $b'(s)$. (Let us recall at this point that we are in the Schr\"odinger picture, where operators are time-independent).  As an illustration consider now the time evolution of CS governed by the simple pseudo-Hermitian Hamiltonian
\beq\lb{p ho}
H_{\rm po}= \omega\left(b^\#(s)b(s) + {\fns \frac{1}{2}}\right),
\eeq
where $b^\#(s) = \eta^{-1}b^\dg(s)\eta$, $\omega\in R_+$. System with Hamiltonian of the type (\ref{p ho}) should be called {\it pseudo-Hermitian oscillator}.  At $s=0$ it coincides with the Hermitian harmonic oscillator of frequency $\omega$. In pseudo-Hermitian mechanics the time evolution of initial $\psi_\alf(x,s)$ and $\vphi_\alf(x,s)$, by definition, is given by
\beq\lb{t2}
\bt{l}
$\dstyle \psi_\alf(x,s,t) = e^{-iHt}\psi_{\alf}(x,s)$,\\[2mm]
$\dstyle \vphi_\alf(x,s,t) = e^{-iH^\dg t}\vphi_{\alf}(x,s)$,
\et
\eeq
For $H = H_{\rm po}$  equations (\ref{t2}) produce
\beq\lb{t3}
\bt{l}
$\dstyle \psi_\alf(x,s,t) = e^{-i\omega t/2}\psi_{\alf(t)}(x,s)$,\\[2mm]
$\dstyle \vphi_\alf(x,s,t) = e^{-i\omega t/2}\vphi_{\alf(t)}(x,s),\quad \alf(t)=\alf e^{-i\omega t}$,
\et
\eeq
which shows that the evolution of CS, governed by the pseudo-Hermitian oscillator Hamiltonian $H_{\rm po}$ is temporally stable. Comparing (\ref{t3}) and (\ref{t}) we see that for the system (\ref{p ho}) the time-evolved CS remain stable with constant $s$, $\alf(t)= e^{-i\omega t}$ and $\chi = -i\omega t/2$.

Finally it is worth noting that the pseudo-Hermitian Hamiltonian $H_{\rm po}$ has {\it unbroken} $PT$-symmetry \cite{Bender1}. Indeed, from (\ref{b(s)2}) and (\ref{b(s)2}) it follows that $PT H_{\rm po}PT = H_{\rm po}$, and from (\ref{psi_n 2}) and (\ref{PQ}) it follows that all eigenstates of $H_{\rm po}$ (and of $H^\dg_{\rm po}$ as well) are eigenvectors of $PT$ with eigenvalue $+1$ or $-1$.

\section*{Conclusion}
We have shown that if the commutator of two non-Hermitian operators $b$ and $\tld b$ equals $1$ and $b$ annihilates a state $\psi_0$ then $\tld b$ is $\eta$-pseudo-Hermitian agjoint $b^\#$ of $b$ and $b^\dg$ is $\eta^{-1}$-pseudo-adjoint of $(\tld{b}^\#)^\dg$. Eigenstates of the pseudo-boson number operator $b^\#b$ and its adjoint $b^\dg (b^\#)^\dg$ form a bi-orthonormal system of pseudo-boson Fock states, while eigenstates of $b$ and its complementary lowering operator $b'=(b^\#)^\dg$ are shown to form {\it bi-normalized and bi-overcomplete} system.This system of states is regarded as system of coherent states (CS) for pseudo-Hermitian bosons. We have provided a simple one-parameter family of ladder operators $b(s)$ and $\tld{b}(s)$ that possess the above described properties and constructed the wave functions of the related Fock states and CS. Fock state wave functions are obtained as product of an exponential of a quadratic form of $x$ and one of the two new polynomials $P_n(x)$, $Q_n(x)$ that are bi-orthogonal and at $s=0$ recover the standard Hermite orthogonal polynomials. The pseudo-boson CS are shown to be temporally stable for the pseudo-boson oscillator Hamiltonian  $\omega\left(b^\#(s)b(s) + 1/2\right)$.

\end{document}